\documentclass[superscriptaddress,twocolumn,showpacs,pra,nofootinbib,10pt]{revtex4-1}
\usepackage{amsmath,xcolor,amsfonts,amsthm,amssymb,graphicx, srcltx,hyperref,ulem}

\usepackage{amsfonts}
\usepackage{amsmath}
\usepackage{epsfig}
\usepackage{xcolor}
\usepackage{ulem} %strikethrough text, call with \sout{arg}

\newcommand{\beq}{\begin{equation}}
\newcommand{\eeq}{\end{equation}}
\newcommand{\ket} [1] {|#1\rangle}

\renewcommand{\emph}[1]{{\it #1}}

\begin{document}

\title{Free-space quantum key distribution by rotation-invariant twisted photons}
\author{Giuseppe Vallone}
\affiliation{Dipartimento di Ingegneria dell'Informazione, Universit\`a di Padova, I-35131 Padova, Italy}
\author{Vincenzo D'Ambrosio}
\affiliation{Dipartimento di Fisica, Sapienza Universit\`a di Roma, I-00185 Roma, Italy}
\author{Anna Sponselli}
\affiliation{Dipartimento di Fisica e Astronomia, Universit\`a di Padova, I-35131 Padova, Italy}
\author{Sergei Slussarenko}
\altaffiliation{Current address: Centre for Quantum Dynamics, Griffith University, Brisbane 4111, Australia}
\author{Lorenzo Marrucci}
\affiliation{Dipartimento di Fisica, Universit\`a di Napoli Federico II and CNR - SPIN, Napoli.}
\author{Fabio Sciarrino}
\affiliation{Dipartimento di Fisica, Sapienza Universit\`a di Roma, I-00185 Roma, Italy}
\affiliation{Istituto Nazionale di Ottica (INO-CNR), Largo E. Fermi 6, I-50125 Firenze, Italy}
\author{Paolo Villoresi}
\affiliation{Dipartimento di Ingegneria dell'Informazione, Universit\`a di Padova, I-35131 Padova, Italy}
\email {paolo.villoresi@dei.unipd.it}
\date{\today}

\begin{abstract}
``Twisted photons'' are photons carrying a well-defined nonzero value of orbital angular momentum (OAM). 
The associated optical wave exhibits a helical shape of the wavefront (hence the name) and an optical vortex at the beam axis.
The OAM of light is attracting a growing interest for its potential in photonic applications 
ranging from particle manipulation, microscopy and nanotechnologies, to fundamental tests of quantum mechanics, classical data multiplexing and quantum communication.
Hitherto, however, all results obtained with optical OAM were limited to laboratory scale. 
Here we report the experimental demonstration of a link for free-space quantum communication with OAM operating over a distance of 210 meters. Our method exploits OAM in combination with optical polarization to encode the information in rotation-invariant photonic states, so as to guarantee full independence of the communication from the local reference frames of the transmitting and receiving units. In particular, we implement quantum key distribution (QKD), 
a protocol exploiting the features of quantum mechanics to guarantee unconditional security in cryptographic communication,
demonstrating error-rate performances that are fully compatible with real-world application requirements. Our results extend previous achievements of OAM-based quantum communication by over two orders of magnitudes in the link scale, providing an important step forward in achieving the vision of a worldwide quantum network.
\end{abstract}

\maketitle

{\it Introduction - }Quantum Key Distribution (QKD) 
has taken its initial steps out of the laboratory years ago and is nowadays included in commercial devices that use 
optical fiber (wireline) systems. With state-of-the-art technology, QKD on a regional scale has been already demonstrated in 
USA, Europe, China, Canada and Japan.
However, free-space optical links are required for long-distance communication among areas which are not suitable for 
fiber installation or for moving terminals, including the important case of satellite-based links. 
Among the various implementation schemes proposed so far, the protocol reported by D'Ambrosio et al. \cite{damb12nco} 
is unique in its exploitation of spatial transverse modes of the optical beam, in particular of the OAM degree of freedom, 
in order to acquire a significant technical advantage, that is the insensitivity of the communication to the relative alignment 
of the users' reference frames. This advantage may be very relevant for the QKD implementations to be upgraded from 
the regional scale to a national or continental one, or for links crossing hostile ground, and even for envisaging a QKD 
on a global scale by exploiting orbiting terminals on a network of satellites.

\begin{figure*}
\includegraphics[width=0.9\textwidth]{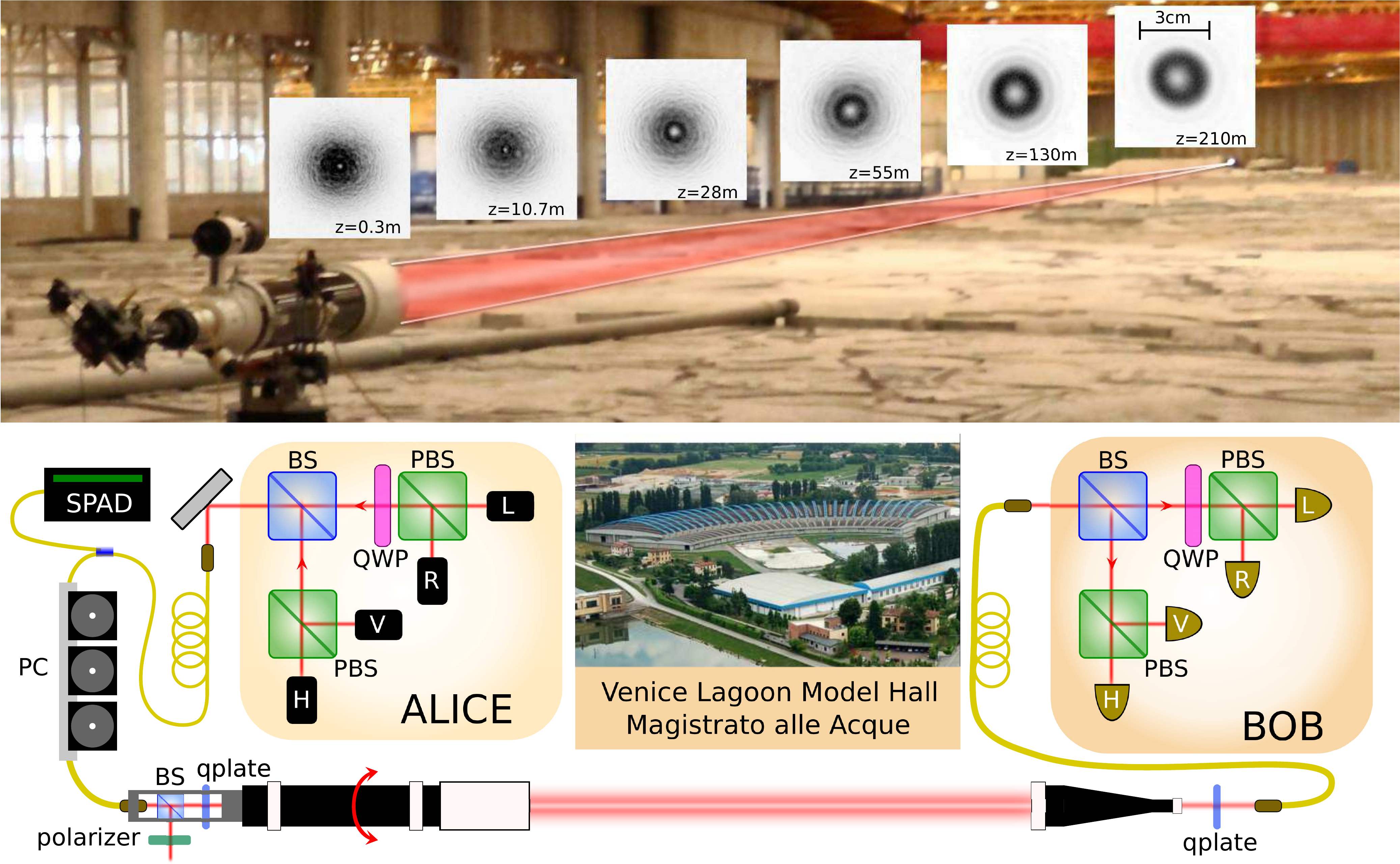} 
\caption{ {\bf Experimental setup}. 
Top image: the experimental layout inside the Venice Lagoon Model Hall located in Padua. Alice's telescope is on the left and Bob's one on the far right. The beam profile versus propagation distance $z$ from the telescope is shown in square frames, all having the same side length of 60 mm. Bottom image: Alice's and Bob's setups. L, R, V and H are four attenuated lasers that generate the polarization qubits, calibrated by using a single photon detector (SPAD). Before starting the QKD, a polarization controller (PC) is used to impose the correct transfer of the polarization states to the telescope by an optical fiber. On the telescope front focal plane, the polarization qubits are transformed by the $q$-plate into hybrid polarization-OAM qubits and sent toward the receiver. At Bob's terminal, the beam is gathered by a 15x beam compressor and sent to the second $q$-plate. Here the hybrid polarization-OAM qubits are transformed back into polarization ones, which are then coupled into a single-mode fiber for spatial-mode filtering and transferred to the BB84 state analyzer.}
\label{setup}
\end{figure*}

In the last decades, the OAM of light has been widely investigated as a new promising resource both for fundamental and applied physics. 
Indeed properties of OAM have been exploited for applications in many different fields, such as microscopy \cite{furh05ope}, astronomy
 \cite{tamb10epl,tamb11nph}, metrology \cite{damb13nco}, fundamental quantum mechanics \cite{damb13prx}, and biophysics \cite{grie03nat}. 
This degree of freedom of light can be used to specify the azimuthal dependence of the transverse modes of the electromagnetic radiation. The OAM eigenmodes are characterized by a twisted wavefront composed of $\ell$ intertwined helices, where $\ell$ is an integer, and by photons carrying $\ell\hbar$ of (orbital) angular momentum, in addition to the more usual spin angular momentum (SAM) associated with polarization. The potentially unlimited value of $\ell$ opens the possibility to exploit OAM also for increasing the capacity of communication systems (although at the expense of increasing also the channel cross-section size), and classical data transmission based on OAM multiplexing has been recently demonstrated both in free-space \cite{gibs04ope,wang12npho,su12ope} and optical fibers \cite{bozi13sci}. Such a feature can also be exploited in the quantum domain, for example to expand the number of qubits per photon \cite{moli07nph,vite13npho,damb13sre}, or to achieve new functions, such as the rotational invariance of the photonic qubits, as in the present work.

OAM is a conserved quantity for light propagation in vacuum, which is obviously important for communication applications. However, OAM is also highly sensitive to atmospheric turbulence, a feature which limits its potential usefulness in many practical cases. Several theoretical studies have been devoted to the investigation of turbulence effects on the OAM propagations in free-space \cite{pate05prl,roux11pra,sanc11ope,mali12ope}. Experiments were performed by simulating the effects of the atmospheric turbulence by adding random phase fluctuations \cite{hama13pra} or by using turbulence cells \cite{pors11ope}. An experiment of classical communication  with free-space propagation of OAM was performed with radio-waves (at a wavelength $\lambda=12.5$ cm) at a propagation distance of $d=442$ m (corresponding to about 3500 wavelengths), in which OAM states were used as a multiplexing resource for classical communication \cite{tamb12njp}. 
In the optical domain, free-space classical communication of OAM has been demonstrated in \cite{gibs04ope}.
 None of these experiments involved quantum communication. To our knowledge, no free-space propagation of OAM-carrying photons in a quantum regime across distances greater than a few meters has been demonstrated up to now. In the present article, we report the realization of free-space alignment-free QKD across air for a distance $d=210$ m (corresponding to $\sim2.5\times10^8$ wavelengths) exploiting OAM in combination with polarization to encode the quantum bits. As compared with the standard polarization-only encoding, our protocol with OAM has the advantage of being frame-independent, thus removing the need for precise alignment of the transmitting and receiving unit frames (although, of course, not the need for accurate beam pointing). On the other hand, compared to OAM-only, the polarization component in our communication scheme guarantees a high robustness against 
spatial perturbations.

{\it Experiment - }In a general bipartite free-space QKD scenario, two users (usually called Alice and Bob) must establish a shared reference frame (SRF) in order to communicate with good fidelity. Indeed the lack of a SRF is equivalent to an unknown relative rotation which introduces noise into the quantum channel, disrupting the communication. When the information is encoded in photon polarization, such a reference frame can be defined by the orientations of Alice's and Bob's ``horizontal'' linear polarization directions. The alignment of these directions needs extra resources and can impose serious obstacles in long distance free space QKD and/or when the misalignment varies in time \cite{aoli07prl,wabn13njp}. As already mentioned, a possible solution to this problem can be achieved by using rotation-invariant photonic states, which remove altogether the need for establishing a SRF \cite{damb12nco}. Such states are obtained as a particular combination of OAM and polarization modes (hybrid states), for which the transformation induced by the misalignment on polarization is exactly balanced by the effect of the same misalignment on spatial modes. 
More in detail, although circular polarization states are eigenstates of reference frame rotation operator, they acquire different phase-shifts 
causing a rotation of linear polarization states. The adoption of hybrid qubits exactly compensate these phase shifts for both circular polarizations
 hence resulting in rotational-invariant photon states \cite{damb12nco,holl11ope}.
Moreover, this rotation-invariant hybrid space can be also regarded as a particular decoherence-free subspace of the four-dimensional OAM-polarization product Hilbert space, insensitive to the decoherence associated with random rotations.

The hybrid states used in our experiment
 can be generated by a particular space-variant birefringent plate having topological charge $q$ at its center, named ``$q$-plate'' \cite{marr06prl,picc10apl,slus11ope}. In particular, a polarized Gaussian beam (having zero OAM) passing through a $q$-plate with $q=1/2$ will undergo the following transformation:
\beq 
\label{hybrid_states}
(\alpha\ket{R}+\beta\ket{L})_\pi\otimes\ket{0}_{\text{O}}
\, \rightarrow \,
\alpha\ket{L}_\pi\otimes\ket{r}_{\text{O}}+\beta\ket{R}_\pi\otimes\ket{l}_{\text{O}}
\eeq
In the previous expression, $\ket{L}_\pi$ and $\ket{R}_\pi$ denote the left and right circular polarization states 
(eigenstates of SAM with eigenvalues $+\hbar$ and $-\hbar$, respectively), $\ket{0}_{\text{O}}$ 
represents the trasverse Gaussian mode with zero OAM and 
$\ket{l}_\text{O}$ and $\ket{r}_{\text{O}}$ the eigenstates of OAM with $|\ell|=1$ and with eigenvalues $+\hbar$ and $-\hbar$, respectively), 
subscripts $\pi$ and  O stand respectively for polarization and orbital angular momentum. The states appearing on the right
hand side of equation \eqref{hybrid_states} are rotation-invariant states \cite{damb12nco}. 
The reverse operation to \eqref{hybrid_states} can be realized by a second $q$-plate with the same $q$.
 In practice, the $q$-plate operates as an interface between the polarization space and the hybrid one, converting 
 qubits from one space to the other and vice versa in a universal (qubit invariant) way. 
 This in turn means that the initial encoding and final decoding of information in our QKD implementation 
 protocol can be conveniently performed in the polarization space, while the transmission is done in the rotation-invariant hybrid space.

Let us now describe the alignment-free free-space QKD experiment. We implemented the so called BB84 protocol 
\cite{benn84ieee} with decoy states \cite{hwan03prl,ma05pra}. The bits are encoded in two mutually unbiases bases 
$\mathbb{Z}=\{\ket{0},\ket{1}\}$ and $\mathbb{X}=\{\ket{+},\ket{-}\}$, where $\ket{0}$ and $\ket{1}$ are two orthogonal 
states spanning the qubit space and $\ket{\pm}=\frac1{\sqrt2}(\ket0\pm\ket1)$. 
Alice randomly chooses between the $\mathbb{Z}$ and $\mathbb{X}$ basis to send the classical bits $0$ and $1$.
 In our hybrid encoding, the $\mathbb Z$ basis states correspond to 
 $\{\ket{L}_\pi\otimes\ket{r}_{\text{O}},\ket{R}_\pi\otimes\ket{l}_{\text{O}}\}$, while the $\mathbb X$ basis states 
 correspond to $\frac1{\sqrt{2}} (\ket{L}_\pi\otimes\ket{r}_{\text{O}} \pm \ket{R}_\pi\otimes\ket{l}_{\text{O}})$. 
 The transmitter uses four different polarized attenuated lasers to generate the quantum bits. 
 Photons were delivered via a single-mode fiber to a telescope. 
 Polarization states $\ket{H}$, $\ket{V}$, $\ket{R}$ and $\ket{L}$ (as defined in the transmitter reference frame) 
 are then transformed into rotation-invariant hybrid states by means of a $q$-plate with $q=1/2$ (see Figure \eqref{setup}). 
 The photons are then transmitted to the receiving station, where a second $q$-plate transforms them back in 
 the original polarization states $\ket{H}$, $\ket{V}$, $\ket{R}$ and $\ket{L}$, as defined in the receiver reference frame. 
 Qubits are then analyzed by 
polarizers and single photon detectors. The sifted key is obtained by keeping only the bits 
corresponding to the same basis at Alice and Bob sides.
{The measured transmission of the channel is $\eta_\text{ch}\sim10\%$ (it includes all the losses in the receiving optics 
before coupling into single mode fiber) while the coupling efficiency into single mode fiber is given by $\eta_\text{c}\sim25\%$--$35\%$.}

To experimentally demonstrate that our apparatus is insensitive to the relative reference frames of the users, 
we performed different QKD experiments by rotating the first telescope around the link axis by the following angles: 
$\theta=15^\circ$, $45^\circ$ and $\theta=60^\circ$. After each rotation, {at Alice's side}, 
we compensated the possible polarization alterations arising from the fiber torsion within the transmitting 
unit and reoptimized the beam pointing. No compensation was performed at Bob's terminal. 
\begin{figure}
\includegraphics[width=8.5cm]{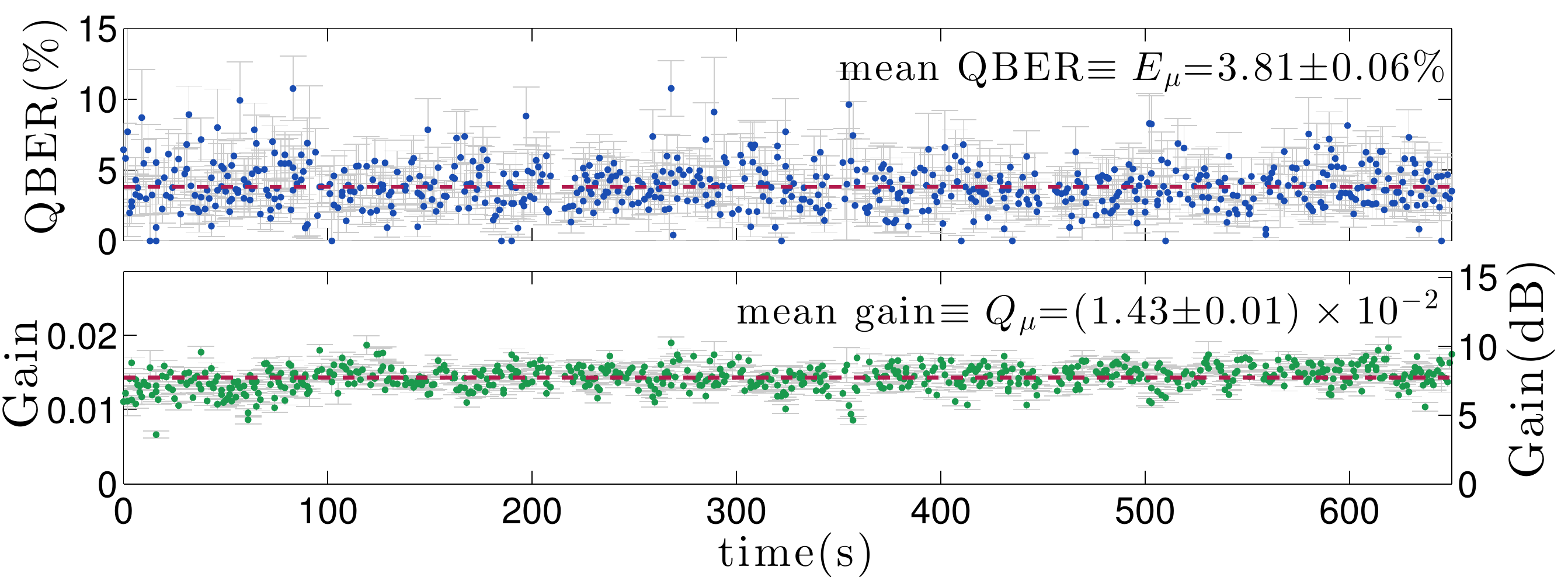} 
\caption{{\bf Experimental QBER and gain}. We show the experimental QBER and gain of the signal state $\mu$ 
{(and their statistical errors)} at zero rotation angle
{for each block of 2880 raw bit data}. The data show 10 minutes of acquisition. Dashed lines represent mean values 
$E_\mu=3.81\%$ and $Q_\mu=1.41\times10^{-2}$ used for the key rate evaluation
while $0.06\%$ and $0.01\times 10^{-2}$ represent the errors on the mean values.
QBER and gain fluctuations from block to block are due to transmission fluctuation caused by the channel turbulence
{and to the finite size of the blocks}.}
\label{gain}
\end{figure}

The experiment was performed at the Experimental Center for Hydraulic Models of the "Magistrato alle Acque di Venezia'', 
inside Venice Lagoon Model Hall (see inset of figure \ref{setup}).
This allow us the demonstration of the polarization-OAM encoding and its decoding by single mode fiber coupling in a free-space
 channel.
The location was chosen in order to study a substantial free-space propagation in reproducible environmental conditions.
The effect of the turbulence on secure key generation 
is analyzed in Appendix \ref{turbo_analysis}.
Alice setup is based on four attenuated lasers ({with wavelength of 850nm}) with four different polarizations: 
horizontal H, vertical V, left-handed L and right-handed R (L and R are obtained by the combination 
of the laser's polarization and a quarter wave plate QWP). 
Qubits are sent in blocks of 2880 bits. An FPGA-based controller drives the four lasers at a frequency of 2.5MHz. 
In our experimental realization, due to constraints with FPGA data communication, 
the sending rate resulted of about 30kbit/s. 
Photons are coupled into a single mode fiber and a fiber beam splitter delivers 95\% of the signal 
to the transmitter telescope and 5\% to a single-photon detector (SPAD) used to estimate the mean photon number per pulse. 
In the telescope cage, a beam splitter (BS) delivers half of the photons to a polarizer, and the other half to a 
$q$-plate with topological charge $q=1/2$. The polarizer, together with the polarization controller (PC), 
allows Alice to correct the fiber-induced polarization rotations. The $q$-plate maps polarization-encoded 
qubits into qubits encoded in hybrid polarization-OAM states. After the $q$-plate, the photons enter a 12 cm 
aperture telescope with focal length $f=900$ mm ($12\times$ of magnification) and are sent toward the receiving station. 
The magnification was chosen in order
to have a nearly collimated beam  between the transmitter and the receiver.
We used a telescope aperture large compared to the beam diameter 
in order to minimize the diffraction effects due to beam clipping.
The telescope is mounted on a rotation stage, allowing the rigid rotation of the entire transmitter device.
 At Bob's receiving station, a 5 cm aperture telescope collects the photons 
 (with losses due to beam clipping of the
 order of 2\%), and a second $q$-plate decodes 
 the hybrid polarization-OAM qubit states into polarization qubits. The photons are then coupled into a single-mode 
 fiber and delivered to the measurement device. 
The four different single-photon detectors are associated to the four polarization states $H$, $V$, $R$ and $L$.

\begin{figure}[t]
\includegraphics[width=7.5cm]{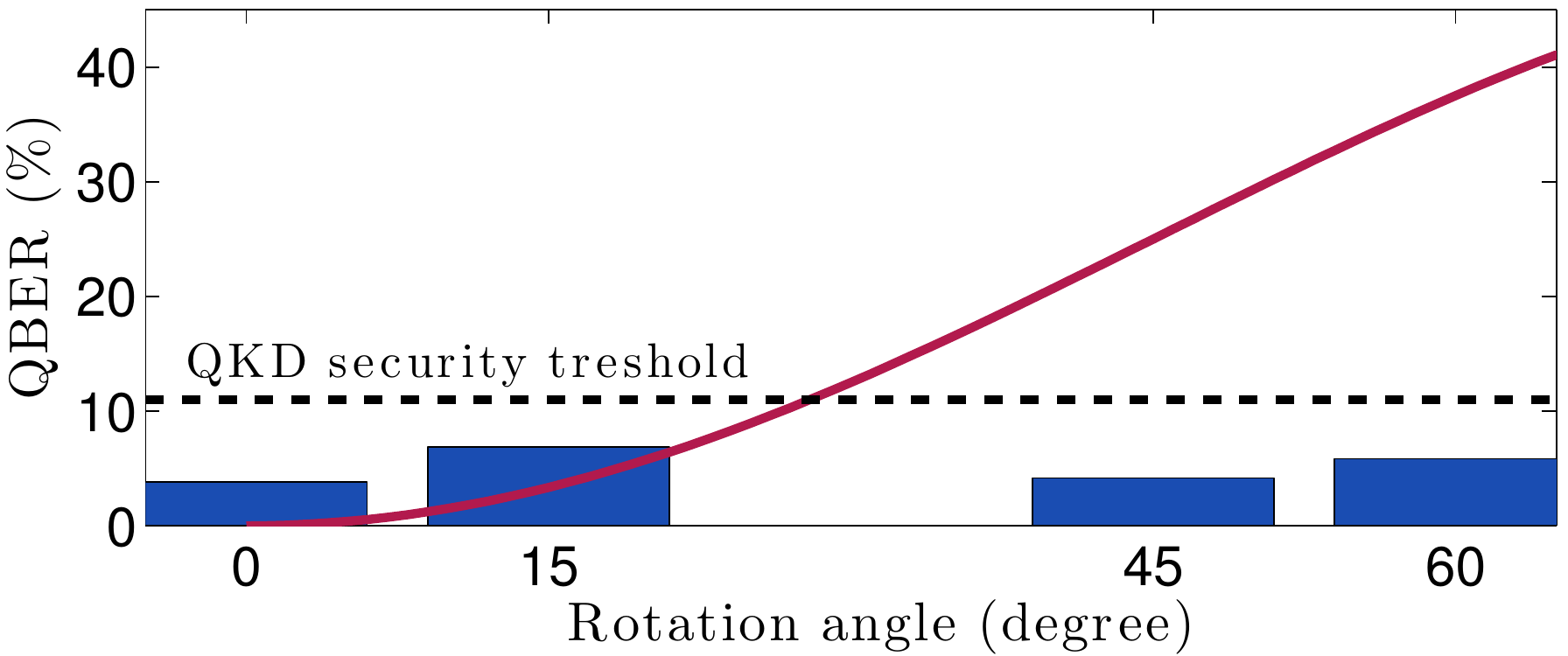}
\caption{ {\bf Mean QBER}  We show the quantum bit error rates (QBERs) obtained for the signal state $\mu$ at four different rotation angles of the transmitter device, $0^\circ$, $15^\circ$, $45^\circ$ and $60^\circ$. 
{The mean QBERs are evaluated over $10^5$ sifted bits and their statistical errors are below $0.1\%$.} 
The dashed line represent the upper limit of the QBER allowing positive key rate in the infinite key limit with true single photon sources, namely $11\%$. The continuous red line represents the theoretical QBER that would be obtained by standard polarization encoding, using the polarization states $\ket{H}$, $\ket{V}$, $\ket{L}$ and $\ket{R}$.}
\label{qberL}
\end{figure}

{\it Secret key rate analysis - }After the sifting phase, the receiver and the transmitter have two correlated bit sequences:
we then measured the quantum bit error rate (QBER) defined as the ratio of the number of errors over the received bits. 
In Figure \ref{gain} we show 10 minutes of transmission: in particular we report the QBER and the gain of the signal 
states for the angle $\theta=0$. 
The gain is the ratio of detected bits over the sent bits. It is possible to see that the transmission is stable over several minutes. 
Next, we {report} the QBER of the protocol for different angles. It is important to notice that, if polarization states were used instead 
of the hybrid rotation-invariant states, the QBER would depend on the relative angle $\theta$ between Alice's and Bob's reference frames as 
QBER$=\frac12\sin^2\theta$ (the circular states $\ket{R}$ and $\ket{L}$ are rotationally invariant, while the linear polarization states 
will introduce an error of $\sin^2\theta$). In Figure \ref{qberL} we show the experimental mean QBER obtained at different 
angles and compare it with the theoretical QBER that would be obtained with polarization encoding. 
It is evident that, as expected, the angle of rotation does not change the QBER: 
the observed variations in the the measured QBER are ascribed mainly to imperfections in Alice's local polarization compensation.

\begin{figure}[t]
\includegraphics[width=7cm]{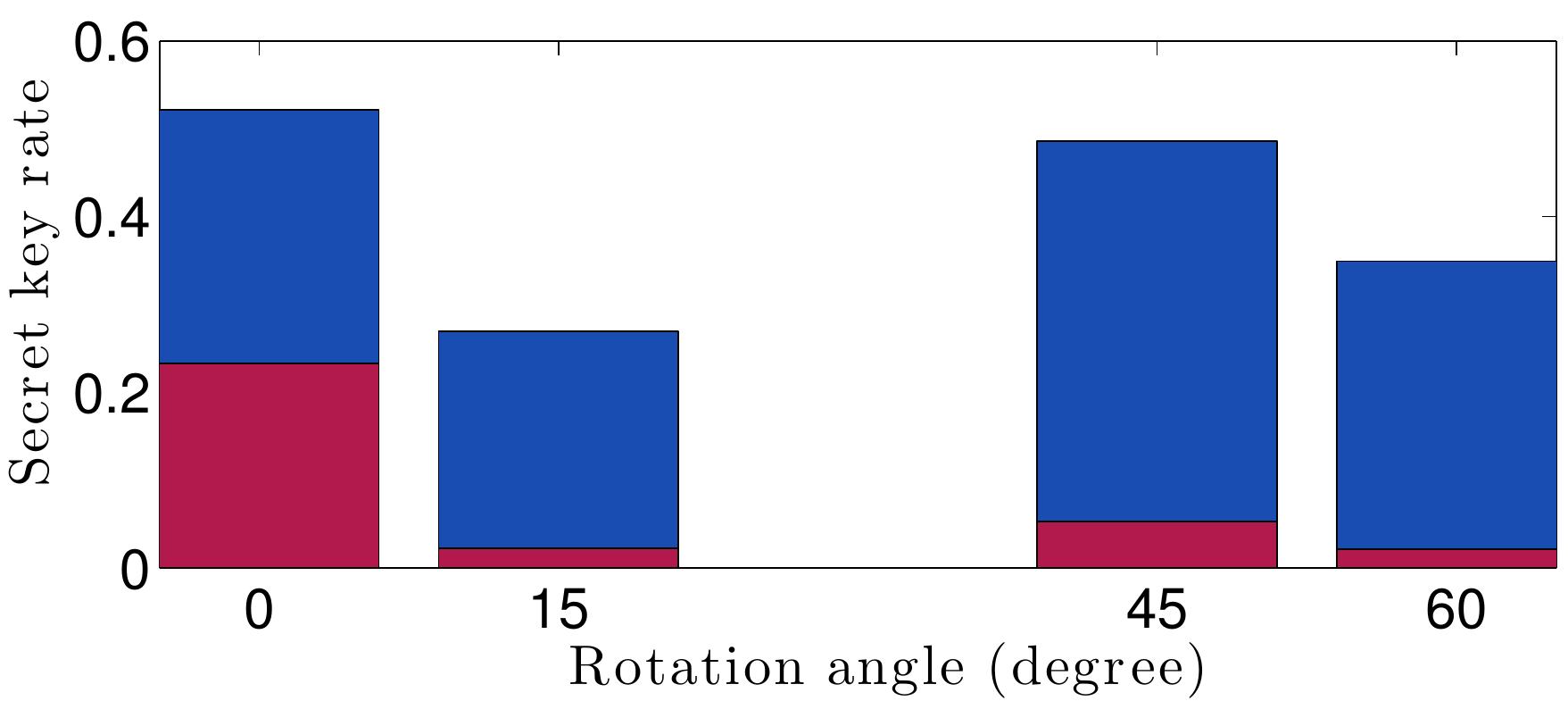}
\caption{ {\bf Secret key rates at different angles} Secret key rates obtained with the decoy state method (red bars) and corresponding key rates achievable with true single photon sources (blu bars).}
\label{qberR}
\end{figure}
Finally, we extracted a secret key between the two users, by using the so called decoy state method \cite{hwan03prl,ma05pra}, avoiding photon 
number splitting attack on the QKD protocol. 
We obtained secret keys according to the
the secret key rate (defined as the number of secure over sifted bits) in the infinite key-length limit given by:
\beq
r=\frac{Q^L_1}{Q_\mu}[1-h_2(e^U_1)]-{\rm leak}_{EC}+\frac{Q_0}{Q_\mu}\,.
\eeq
The term ${\rm leak}_{EC}$ represent the fraction
of revealed bits during the {performed} classical error correction based on the Cascade protocol, whose efficiency, given by 
$f(E_\mu)=\frac{{\rm leak}_{EC}}{h_2(E_\mu)}$, is below $1.05$. 
See Appendix \ref{data} for the definition and the values measured  for different rotation angles 
 of the different terms appearing in the previous equation.

In figure fig. \ref{qberR}  we show the achieved secret key rates for different angles. 
We also compare the rates obtained with the decoy state method with the optimal rates that could be achieved 
with communication performed at the same QBER but using a true single-photon source at the transmitter.
In this case the achievable key rate becomes $r=1-h_2(E_\mu)-{\rm leak}_{EC}$. 
In Appendix \ref{secure_distance} we also give an estimation of the maximum distance for the secure key generation
 in our conditions.

{\it Conclusions - }In summary, 
we have demonstrated free-space QKD across 210 m without need for aligning the reference frames of the transmitting and receiving units. This was achieved by using rotation-invariant photonic states obtained as specific polarization-OAM combinations. The demonstrated method is in principle not limited to QKD, but could be extended to full quantum communication of arbitrary qubits. 
Our experiment provides also a new record of distance for free-space quantum communication with OAM states, extending previous achievements by more than two orders of magnitude. The extension of the present results to larger Hilber space is left for future research program. 

Note that, after the completion of the present
manuscript, classical communication of different OAM states has been
reported in \cite{kren14qph}.

\acknowledgements
We would like to thank Nicola Laurenti and Matteo Canale of the University of Padova for useful discussions.
Our work was supported by the Strategic-Research-Project QUINTET of the Department of Information Engineering, University of Padova, the Strategic-Research-Project QUANTUMFUTURE of the University of Padova,
Progetto di Ateneo PRAT 2013 ``OAM in free space: a new resource for QKD'' (CPDA138592),
FIRB-Futuro in Ricerca HYTEQ and QWAD (Quantum Waveguides Applications \& Development). We would like to thank also the ''Magistrato alle Acque di Venezia'', for granting us access to the Venice Lagoon Model Hall at the Experimental Center for Hydraulic Models in Padua and the ''Ottica San Marco'' for giving us the ''SkyWatcher'' telescope used in the experiment.

\newpage
\onecolumngrid
\appendix
\section{Experimental data used for key extraction}
\label{data}
In this section we report the experimental data used to extract the secure key in the QKD protocol.
Decoy state method has been developed \cite{hwan03prl,ma05pra} 
to avoid photon number splitting attacks on qubits generated by attenuated laser pulses. 
The transmitter randomly changes the mean photon number of the sent pulses between three values: 
$\mu$, the signal state, and two other decoy values, $\nu$ and zero (corresponding to sending empty pulses). 
The bits obtained with $\mu$ are used to build the final key, while other pulses are used to bound 
Eve's knowledge on the key.

In the infinite key-length limit, the secret key rate, defined as the number of secure over sifted bits, is given by \cite{hwan03prl,ma05pra} 
\beq
\label{key_rate}
r=\frac{Q^L_1}{Q_\mu}[1-h_2(e^U_1)]-{\rm leak}_{EC}+\frac{Q_0}{Q_\mu}
\eeq
where $Q_\mu$ is the total gain (the fraction of detected bits over the sent bits), $Q^L_1$ the lower bound of the gain of the one-photon states, 
{$Q_0$ the gain of the vacuum states,  $E_\mu$ the total quantum bit error rate (QBER), $e^U_1$ the upper bound of errors of the one-photon states, $h_2$ the binary entropy $h_2(x)=-x\log_2x-(1-x)\log_2(1-x)$. The term ${\rm leak}_{EC}$ represent the fraction
of revealed bits during the classical error correction protocol, whose efficiency, given by $f(E_\mu)=\frac{{\rm leak}_{EC}}{h_2(E_\mu)}$, 
is below $1.05$}. 

By the decoy state method it is possible to estimate the parameters $Q^L_1$, $Q_0$
 and $e^U_1$ by the decoy data as
\beq
\label{Q1}
Q^L_1=\frac{\mu^2e^{-\mu}}{\mu\nu-\nu^2}\left({Q_\nu}e^\nu-{Q_\mu}e^\mu\frac{\nu^2}{\mu^2}-\frac{\mu^2-\nu^2}{\mu^2}Y_0\right)
\eeq
and
\beq
\label{e1}
e^U_1=\frac{E_\nu Q_\nu e^  \nu-e_0Y_0}{Q^L_1\frac{\nu}{\mu}e^\mu}\,,\quad Q_0=e^{-\mu}Y_0\,.
\eeq
In the previous expression $Q_\nu$ and $E_\nu$ are the total gain and the QBER of decoy states $\nu$ respectively,
{while $Y_0$ is the dark rate at the receiver}. 
The parameters $\mu$ and $\nu$ represent the measured value of signal and decoy mean photon number per pulse 
at the transmitter, respectively given by $\mu=0.623\pm0.002$ and $\nu=0.165\pm0.001$.

\begin{table}[b!]
\begin{center}
\begin{tabular}{c|c|c|c|c|c|}
\hline
  $\theta$   & $Q_\mu$ & $E_\mu (\%)$ & $Q_\nu$ & $E_\nu(\%)$ & $Y_0$\\
\hline
$0^\circ$ & $1.43\times10^{-2}$ & $3.81$   & $4.77\times10^{-3}$    & $7.63$    &$3.77\times10^{-4}$
\\
$15^\circ$ & $1.30\times10^{-2}$ & $6.88$  & $4.12\times10^{-3}$    &  $8.67$  &$2.55\times10^{-4}$
\\
$45^\circ$ & $1.11\times10^{-2}$ & $4.16$    & $2.77\times10^{-3}$    &  $4.47$  &$6.63\times10^{-5}$
\\
$60^\circ$ & $0.85\times10^{-2}$ & $5.84$  & $2.34\times10^{-3}$    &  $6.23$  &$1.13\times10^{-4}$
\\
\hline
\end{tabular}
\end{center}
\caption{Experimental values of the signal and decoy gains, $Q_\mu$ and $Q_\nu$ and corresponding QBER $E_\mu$ and $E_\nu$ for
different rotation angles $\theta$. We also report the background rate $Y_0$.}
\label{table}
\end{table}

In table \ref{table} we show the experimental measured parameters $Q_\mu$, $E_\mu$, $Q_\nu$,  $E_\nu$ and $Y_0$
used to estimate the secure key rate.
 The measured $Q_\mu$ are compatible with the the measured transmission of the channel $\eta_\text{ch}\sim10\%$,
  the coupling efficiency into single mode fiber $\eta_\text{c}\sim25\%$--$35\%$,
   the detection efficiency $\eta_\text{d}\sim60\%$ since $Q_\mu\simeq\mu\eta_\text{ch}\eta_\text{c}\eta_\text{d}$.

\section{Analysis of the turbulence}
\label{turbo_analysis}
\begin{figure}[h]
\centering\includegraphics[width=18cm]{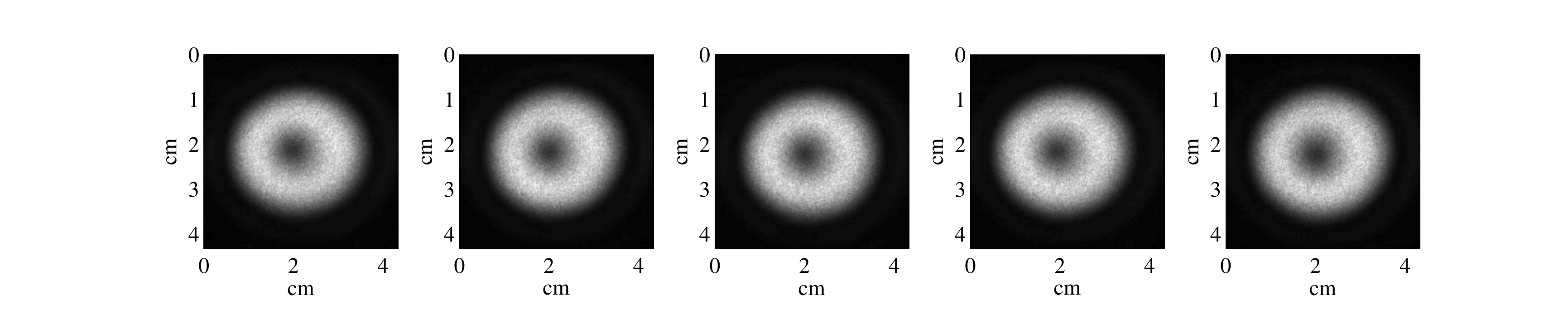}
\caption{Intensity pattern recorded at the receiver at different times. The delay between images is 10s.}
\label{OAM_intensity}
\end{figure}

In this section we investigate the effects of the turbulence on the OAM propagation.
We recorded the intensity pattern at the receiver for 30 seconds at a frame rate of 4.95fps, obtaining 177 frames.
We show in figure \ref{OAM_intensity} some of the recorded frames. 

\begin{figure}[h!]
\centering\includegraphics[width=18cm]{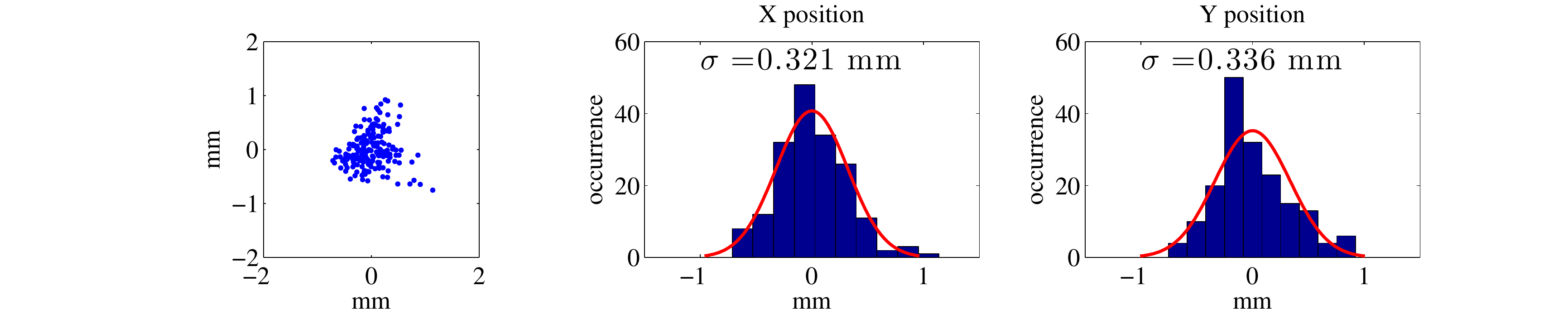}
\caption{Centroids position and their distributions in the $X$ and $Y$ axis.}
\label{centroids}
\end{figure}

In case of weak turbulence, the main effect on the beam propagation is the so called beam wandering,
corresponding to a movement of the intensity centroids at the receiver plane.
In order to evaluate the turbulence parameters, we calculated the centroid positions in each frame. Their positions and they corresponding
distributions in the $X$ and $Y$ axis are reported in fig. \ref{centroids}.

In weak turbulence condition, the standard deviation $\sigma_m$ of the displacement of the centroids
is  related to the Fried parameter (or Fried's coherence length)
$r_0$ according to the relation given by Fante \cite{fant75ieee}:
\beq
\sigma^2_m=\frac{4L^2}{k^2r^2_0}
\eeq
where $L$ is the path length and $k=2\pi/\lambda$ the wavevector of the optical beam.  
We extended the previous equation for OAM beam since the turbulence is weak and its main effect on the
beam propagation is beam wandering. 

In our case, the measured average value is $\sigma_m=0.33mm$, and the corresponding estimate for $r_0$ is
\beq
r_0=\frac{2L}{k\sigma_m}\simeq 17cm\,.
\eeq
Since the beam radius $r\simeq1.5cm$ is much smaller than $r_0$
the effects of the phase aberrations are weak and the OAM scattering is small \cite{pate05prl}.
 Indeed, the chief ray of the beam is moving of the order of a millimeter in the transverse plane of the receiver, with a limited increase of losses.
In these conditions we can also exclude the generation of OAM states 
which is predicted for stronger turbulence in \cite{oesc13ope} (see also references therein and \cite{mali12ope}).

The Fried parameter can be related to the 
 atmospheric turbulence strength $C^2_n$ by the following equation
\beq
r_0 = [ 0.423 k^2 \int C_n^2(z) dz  ]^{-3/5} \,.
\eeq
By assuming that the $C^2_n$ coefficient is constant along the path $C^2_n(z)=C^2_n$ we can obtain
\beq
C^2_n=\frac{r_0^{-5/3}}{0.432\, k^2\, L}\simeq 4 \cdot 10^{-15} m^{-2/3}\,,
\eeq
indeed corresponding to weak turbulence.

\section{Secure distance analysis}
\label{secure_distance}
By using the data of our experiment we can estimate the maximum distance for the secure key generation. 
We consider the dark count of the (free-running) detectors equal to 100Hz (a typical condition achievable in dark condition).
Since the duration of our qubits is 50ns, it is possible to estimate the dark rate as $Y_0=5\cdot 10^{-6}$.
By considering a typical channel QBER of $E_{\rm ch}=2\%$, we can predict the error rate in
the signal and decoy transmission as
\beq
E^*_\mu=\frac12\frac{Y_0}{Q_\mu}+E_{\rm ch}(1-\frac{Y_0}{Q_\mu})\,,
\qquad
E^*_\nu=\frac12\frac{Y_0}{Q_\nu}+E_{\rm ch}(1-\frac{Y_0}{Q_\nu})
\eeq
with $Q_\nu=\frac{\nu}{\mu}Q_\mu$. We here remember that, due to
our polarization-OAM encoding, turbulence will affect the losses and not the QBER.

\begin{figure}[h]
\centering\includegraphics[width=12cm]{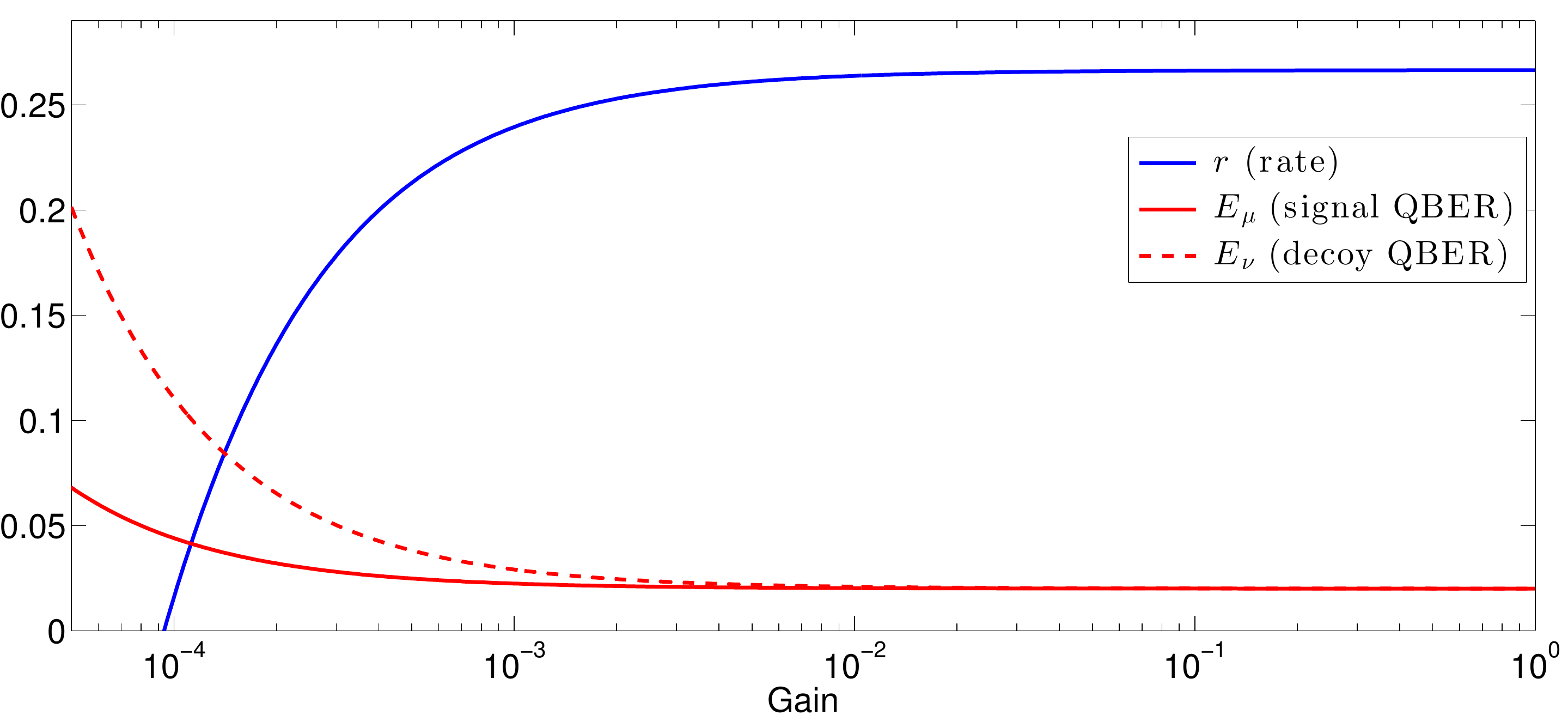}
\caption{Expected rate and QBER in function of the gain $Q_\mu$.}
\label{rate_distance}
\end{figure}

By using equations \eqref{key_rate} with $f(E_\mu)=1.05$, $\mu=0.623$ and $\nu=0.165$, we can estimate the QBERs
$E_\mu$, $E_\nu$ and the key rate $r$ in function
of the signal gain $Q_{\mu}$. The result are shown in fig. \ref{rate_distance}.
Positive rates are obtained up to a gain of $G^*_\mu\simeq 10^{-4}$. 
Since we measured a gain of $G_\mu=1.2\cdot 10^{-2}$, our system could tolerate losses that are two order of magnitude larger.

We estimate that positive rate could be achieved up to few kilometers. Indeed, by using a suitable collecting telescope 
(with diameter of the order of 30cm) it is possible to reduce the losses due to beam clipping in the few $Km$ scale. 
Concerning turbulence effects, with the  atmospheric turbulence strength equal to the value we measured
$C^2_n\simeq 4 \cdot 10^{-15} m^{-2/3}$, the Fried parameter $r_0$ becomes of the order of the beam radius for distance larger
than $1Km$. In this case, as predicted by Paterson \cite{pate05prl}, the scattering between OAM modes become influent and this 
translates for our encoding into additional losses lowering the transmission by a factor of $0.1$. Longer links will produce larger losses
according to \cite{pate05prl}.

\end{document}